\documentclass[12pt]{article}
\usepackage{epsfig}
  
\def\be{\begin{equation}}
\def\ee{\end{equation}}
\def\ben{\begin{displaymath}}
\def\een{\end{displaymath}}
\def\ba{\begin{array}{c}}
\def\bal{\begin{array}{l}}
\def\ea{\end{array}}
\def\p{\partial}
\begin{document}

 \begin{center}
{\tiny .}

\vspace{.35cm}

{\Large \bf

Tridiagonal ${\cal PT}-$symmetric $N$ by $N$ Hamiltonians  and a
fine-tuning of their observability domains

in the strongly non-Hermitian regime.

   }\end{center}

\vspace{10mm}

 \begin{center}

 {\bf Miloslav Znojil}

 \vspace{3mm}
Nuclear Physics Institute ASCR,

 250 68 \v{R}e\v{z}, Czech Republic

{e-mail: znojil@ujf.cas.cz}

\vspace{3mm}

\vspace{5mm}
%

\end{center}

\vspace{5mm}

\section*{Abstract}

A generic ${\cal PT}-$symmetric Hamiltonian is assumed
tridiagonalized and truncated to $N<\infty$ dimensions, $H \to
H^{(chain\ model)}$, and all its up-down symmetrized special cases
with $J=[N/2]$ real couplings are considered, $H^{(chain\ model)}
\to H^{(N)}$. Using symbolic manipulation and extrapolation
techniques we find out that in the strongly non-Hermitian regime
the secular equation gets partially factorized at all $N$. This
enables us to reveal a fine-tuned alignment of the dominant
couplings implying an asymptotically sharply spiked shape of the
boundary of the $J-$dimensional quasi-Hermiticity domain ${\cal
D}^{(N)}$ in which all the spectrum of energies $E_n^{(N)}$
remains real and observable.

\newpage

\section{\label{par1} Introduction}

Unexpectedly often, many general theoretical considerations as
well as practical phenomenological applications of quantum
mechanics rely on the exceptionally friendly mathematics connected
with the one-dimensional harmonic oscillator $H^{(HO)}$. In
particular, the equidistance of its energies (say,
$E=1,3,5,\ldots$ in suitable units) proves fairly favorable in
perturbation theory where it simplifies the practical calculations
of the spectra of the various anharmonic oscillators
 \be
 H^{(AHO)}=H^{(HO)}+
 H^{(perturbation)}\,.
 \label{LAHO}
 \ee
Even when one restricts attention to the mere {\em
finite-dimensional} perturbations written in terms of the
eigenvectors $|\,m^{(HO)}\rangle$ of $H^{(HO)}$ itself,
 \be
 H^{(perturbation)}=
 \sum_{\tiny \ba
 m,n=1
 \ea}^N\,|\,m^{(HO)}\rangle\, \tilde{W}^{(N)}_{m,n}\,\langle
 n^{(HO)}\,|\,,
 \label{AHOfr}
 \ee
the finite-dimensional matrix example (\ref{LAHO}) +
(\ref{AHOfr}) may produce a lot of theoretical inspiration as
sampled, e.g., in Chapter Two of the Kato's classical book on
perturbation theory where many low-dimensional Hermitian as well
as non-Hermitian sample matrices were considered \cite{Kato}.

The parallel, purely phenomenological inspiration by the truncated
eq.~(\ref{AHOfr}) need not be less exciting. For example, the
authors of an explicit numerical exercise \cite{recenze} felt
inspired by the recent growth of popularity of the ${\cal
PT}-$symmetric \cite{Carl} and pseudo-Hermitian \cite{Ali} models
and analyzed their {\em strongly-perturbed} sample with
 \be
  \tilde{W}^{(N)}_{m,n}= {\rm i}\,\langle
 m^{(HO)}\,|\,x^3\exp\left [-\alpha
 \,H^{(HO)}\,
 \right ]
 \,|\,n^{(HO)}\rangle\,, \ \ \ \ m,n=1,2,\ldots,N.
 \label{aryeh}
 \ee
They found out that as long as $H^{(HO)}=x^2+p^2$ is positive, a
very good approximation of the nonlinear-differential-equation
$N=\infty$ results is achieved via the series of truncated, purely
algebraic and linear $N<\infty$ eqs.~(\ref{AHOfr}).

In our recent papers \cite{I} and \cite{condit} we made one more
step. Assuming that one could annihilate {\em any}
far-off-diagonal element of {\em any} given matrix ${H}^{(AHO)}$
via a finite sequence of elementary Jacobi rotations
\cite{Wilkinson} we restricted our attention to the mere
``irreducible", {\em tridiagonal} anharmonic-oscillator-type
models
 \be
 H^{(chain)}=H^{(HO)}+
 \sum_{\tiny \ba
 m,n=1\\
 |m-n|=1
 \ea}^N\,|\,m^{(HO)}\rangle\, W^{(N)}_{m,n}\,\langle
 n^{(HO)}\,|\,.
 \label{AHO}
 \ee
In the ${\cal PT}-$symmetric scenario (guaranteed, in the
normalization accepted in \cite{I}, by the mere antisymmetry
$W^{(N)}_{m,m+1}=- W^{(N)}_{m+1,m}\equiv g_m$ among the {\em real}
matrix elements) we shifted the origin of the energy scale and
arrived at the $N-$dimensional and tridiagonal ``chain-model"
matrices $\langle m^{(HO)}\,|\,H^{(chain)}\,|\,n^{(HO)}\rangle
\,\equiv\, H^{(chain)}_{m,n}$,
  \be
 H^{(chain)}
 =\left [\begin {array}{cccccc}
  1-N&g_1&0&0&\ldots&0\\
 -g_1& 3-N&g_{2}&0&\ldots&0\\
 0&-g_{2}&5-N&\ddots&\ddots&\vdots
 \\
 0&0&-g_{3}&\ddots&g_{N-2}&0
 \\
 \vdots&\vdots&\ddots&\ddots&N-3&g_{N-1}\\
 0&0&\ldots&0&-g_{N-1}&N-1
 \end {array}\right ]\,\neq \,\left (H^{(chain)}\right )^\dagger\,.
 \label{NNIPTS}
 \ee
For the sake of simplicity we decided to pay attention solely to
the up-down-symmetric special cases $H^{(N)}$ of $H^{(chain)}$
where we choose
 \be
 g_{N-k}=g_{k}\geq 0\,,\ \ \ \ \ \ \ k = 1, 2, \ldots, J
 \label{toy}
 \ee
at both the even $N=2J$ and the odd $N=2J+1$.

In the subsection \ref{art1} of section \ref{art} below we shall
start our present continuation of the latter study of the toy
model (\ref{NNIPTS}) + (\ref{toy}) by a brief review of the
results of paper~\cite{I}. We point out there that the model
$H^{(N)}$ itself has been introduced as partially tractable by an
algebraic, symbolic-manipulation-based non-numerical extrapolation
method. We remind the readers that at any dimension $N=2J$ or
$N=2J+1$, the spectrum $\{E_n^{(N)}\}$ remains real and observable
inside a $J-$dimensional domain ${\cal D}={\cal D}^{(N)}$ of the
matrix elements which is compact and {all contained} inside a
bigger domain ${\cal S^{(N)}}$ defined by the following elementary
inequality \cite{I},
 \be
 \frac{N^3-N}{2} \ \geq \ 2\,
 \sum_{n=1}^{J-1}\,g_n^2
 +\left \{
 \begin{array}{ll}
 g_J^2,&N=2J,\\
 2\,g_J^2,\ \ \ \ \ \ \ \ &N=2J+1\,.
 \ea\right .
 \label{righ}
  \ee
There exists just a finite set of the ``maximal-coupling"
intersections of the two surfaces, i.e., of the
$(J-1)-$dimensional boundaries $\p {\cal D}^{(N)}$ and $\p {\cal
S}^{(N)}$. We succeeded in determining the coordinates of these
points (called, in ref.~\cite{I}, extremely exceptional points,
EEP) in closed form,
 \be
 g_n^{(EEP)} = \sqrt{n\,(N-n)}\,, \ \ \ \ \ n = 1, 2, \ldots, J\,.
 \label{optim}
 \ee
In the subsection \ref{art2} we shall extend the review by adding
some empirical observations published in our most recent numerical
study \cite{condit} and concerning the behaviour of the energies
$E_n^{(N)}$, predominantly, {\em far off} the EEP extremes. In
particular, we shall recollect there the lucky guess of the ansatz
 \be
 g_n= g_n^{(EEP)} \,\sqrt{\left(1-\xi_n(t) \right) }\,,
 \ \ \ \ \ \ \ \ \
 \xi_n(t) = t+t^2+\ldots+t^{J-1}+G_n t^J\,
 \label{lobkov}
 \ee
which extrapolates, to all $J$, the rigorous fine-tuning rule
derived, in ref.~\cite{I}, at $J=2$. In the numerical context of
ref.~\cite{condit} it merely served as a bookkeeping tool in our
experiments with the various choices of the rescaled couplings
$G_n$. In what follows we intend to describe several much deeper
and more far-reaching consequences of this type of an ansatz.

In the preliminary steps made in section \ref{parag3} we shall
start the analysis of our bound-state problem by the direct,
brute-force algebraic solution of its secular equations
 \be
 \det \left ( H^{(N)}-E\,I^{(N)}\right )=0\,
 \label{seculare}
 \ee
at $N=4$ (subsection \ref{par2.3}), at $N=5$ (subsection
\ref{par2.5}) and at $N=6$ (subsection \ref{parag3.3}). In the
subsequent section \ref{parag4} we shall change the variables in
the manner prescribed by eq.~(\ref{lobkov}). This will lead to the
much more compact strong-coupling leading-order  formulae at the
two sample dimensions $N=4$ (subsection \ref{par2.4}) and $N=5$
(subsection \ref{par5.1}). Certain indications of the possibility
of a successful extrapolation of these  formulae with respect to
the dimension $N$ will follow in section \ref{par} sampling $N=6$
(subsection \ref{par4.2}) and $N=7$ (subsection \ref{par5.2}).

The climax of our present paper comes in sections \ref{par4}
(where we extend the above results to all the even dimensions
$N=2J$) and \ref{par5} (where the parallel extrapolations are
outlined in the case of any odd $N=2J+1$). The remaining text are
just discussions (section \ref{parag8}) and summary (section
\ref{parsum}) which emphasize that our present,
perturbation-theory-simulating attention paid to the interval of
{\em small} $t$ fills in fact the gap between the algebraic $t=0$
approach of paper \cite{I} and its numerical large$-t$ pendant of
letter~\cite{condit}.

\section{A brief review of the state of art \label{art} }

Our continuing interest in the family (\ref{NNIPTS}) has several
reasons. Firstly, in possible applications, special cases of
$H^{(N)}$ could play the role of a non-standard spin model (cf.
their $N=2$ samples in refs.~\cite{turek,PLB2}) or of a
Hamiltonian of a system where a ``de-frozen", new degree of
freedom can emerge (with $N=3$ see \cite{PLB3}). At any higher
dimension $N$, every $H^{(N)}$ is, using the language of the
review paper \cite{Geyer}, quasi-Hermitian and, therefore,
eligible, say, as a Hamiltonian of a quantum chain. Inside ${\cal
D}^{(N)}$ {\em and} in a suitably specified physical Hilbert space
${\cal H}^{(N)}$, all our models $H^{(N)}$ obey all the postulates
of Quantum Mechanics in a way illustrated, at $N=2$, in
\cite{PLB2}. Due to their finite matrix form, all of them appear
particularly suitable for an illustration of some subtleties of
perturbation theory, especially in the weak-coupling dynamical
regime where all their off-diagonal elements $g_n$ remain small
\cite{Kato}. In the sense explained in \cite{I}, all of our toy
Hamiltonians $H^{(N)}$ are also interesting as ${\cal
PT}-$symmetric \cite{Carl,BB} and/or parity-pseudo-Hermitian
\cite{Ali} candidates for operators of observables which are
capable of exhibiting multiple confluences of their (in the Kato's
language) ``real exceptional points" \cite{condit,Uwe,Dieter}.
Last but not least, our non-Hermitian matrices $H^{(N)}$ are,
through the above-mentioned variational and perturbative
considerations \cite{kde}, directly and closely related to the
popular {\em differential} ${\cal PT}-$symmetric quantum
Hamiltonians of Bender et al \cite{BB} et al \cite{Caliceti} --
\cite{broken}.

\subsection{Observations made in paper \cite{I} \label{art1} }

In ``paper I" \cite{I} we felt inspired by the exceptional
transparency of the geometry of the ``physical" domains of
quasi-Hermiticity ${\cal D}^{(N)}$ in the strong-coupling regime
and at the smallest dimensions $N$. In such a setting, the first
non-numerical result of ref.~\cite{I} was that {\em all} the
eigenenergies become complex whenever the anharmonicity becomes
sufficiently strong. Such a type of observation was interpreted as
very important because the crucial point of making {\em any}
${\cal PT}-$symmetric Hamiltonian $H$ ``physical" (i.e.,
responsible for observable and stable bound states) lies in the
specification of the ``allowed" range ${\cal D}={\cal D}(H)$ of
its free parameters.

In paper I we showed that the {\em non-numerical} construction of
${\cal D}(H)$ proves feasible in an EEP, ``maximal-coupling" limit
$H=H^{(N)}_{(EEP)}$ of our tridiagonal matrices $H^{(N)}$ at {\em
all} their dimensions $N$. The proof was based on an elementary
observation that for every individual model $H^{(N)}$ with a fixed
dimension $N$, the spectrum is determined by the polynomial
secular equation for the squared energy $s=E^2$,
 \ben
  \det \left [ H^{(N)}-E\,I^{(N)}\right ]=
  s^J-P_1^{(N)}(g_1, \ldots, g_J)\,s^{J-1}
 +P_2^{(N)}(g_1, \ldots, g_J)\,s^{J-2}
 -
 \ \ \ \ \ \ \ \ \ \ \ \ \ \ \ \ \ \
 \een
 \be
 \ \ \ \ \ \ \ \ \ \ \ \ \ \ \ \ \ \
 - \ldots +(-1)^J\, P_J^{(N)}(g_1, \ldots, g_J)=0\,.
 \label{secu}
 \ee
We recollected that due to the polynomiality of the latter
equation, the sum of the {\em physical} (i.e., nonnegative) roots
$s_j$ must be equal to the first coefficient,
 \be
 P_1^{(N)}(g_1,\ldots, g_J)=s_1+s_2+\ldots+s_J\geq 0\,.
 \label{upperes}
 \ee
As a consequence, the closure of the domain ${\cal D}^{(N)}$ must
lie {\em inside} the closure of another, bigger domain ${\cal
S}^{(N)}$ which is defined, much more simply, by the upper
estimate (\ref{upperes}). The shape of the surface of ${\cal
S}^{(N)}$ is a hyperellipsoid or a hypersphere (cf. \cite{I} or
eq.~(\ref{righ}) above). This enabled us to define the EEP
vortices as the points where the couplings are maximal, i.e.,
where the boundary $\p {\cal D}^{(N)}$  of the quasi-Hermiticity
domain {\em intersects} the circumscribed ``upper-estimate"
hypersurface $\p {\cal S}^{(N)}$. In this context, a key technical
result of paper I consisted in the derivation of {\em  the closed
formula} for {all} of the EEP coordinates (viz., of
eq.~(\ref{optim}) above).

In the present continuation of paper I we intend to pay attention
to the shape of the hypersurfaces $\p {\cal D}^{(N)}$ in the
vicinity of their EEP extremes. This is a well motivated project
since, in spite of the reality (i.e., ``mathematical"
observability) of the energies of the system in its solvable EEP
limit, the ``physical" version of the same (i.e., strong-coupling)
observability concept requires more than that. Obviously, during
any measurement we should stay {\em in the interior} of the domain
${\cal D}^{(N)}$, requiring that a small random perturbation of
the couplings {\em cannot} induce a spontaneous complexification
of some energies and a subsequent sudden collapse of the system.

\subsection{Observations made in paper \cite{condit} \label{art2} }

Later on, we complemented the algebraic constructions of
ref.~\cite{I} by the purely numerical study \cite{condit} of the
possible complexification patterns of the spectra of our chain
models $H^{(chain)}$. We can summarize that

 \begin{itemize}

 \item
the perturbed harmonic-oscillator spectrum $\{E_n\}$ remains all
real in the weakly anharmonic regime characterized by a
``sufficient" smallness of all the elements or couplings
$H^{(N)}_{m,m+1}$ at all $m$;

\item the pairs $(E_n,E_{n+1})$ of energies coincide and
complexify whenever their mutual coupling $H^{(N)}_{n,n+1}$
exceeds certain  $n-$dependent ``exceptional-point" value
\cite{Kato}.

\end{itemize}

 \noindent
A particularly amusing empirical (though easily explained)
observation made in \cite{condit} was that one can preserve the
reality of the spectrum even when crossing  the no-interaction
boundary in accordance with the rule
 \be
 g_k \sim \sqrt{\lambda}\,,
 \ \ \ (\lambda > 0) \ \ \longrightarrow \ \  (\lambda < 0)\,.
 \label{lamb}
 \ee
Although such a switch between the {\em real} and {\em purely
imaginary} couplings will not be considered in what follows, we
shall still use the parametrization (\ref{lamb}) in its
strong-coupling real form (\ref{lobkov}) where $\lambda = 1-t$ is
assumed large while the new, formal auxiliary parameter $t \in
(0,1)$ remains, preferably, sufficiently small, characterizing our
${\cal PT}-$symmetric and up/down symmetric chain models $H^{(N)}$
in their most interesting, strongly non-Hermitian dynamical
regime.

In \cite{condit}, the use of the parameters $\lambda$ or $t$ has
been shown to facilitate the study of the spectra, real or
complex, at any $N$. In addition, the use of the ``renormalized"
coupling strengths $G_n$ opened the way towards a systematic
(viz., combinatorial) classification of the non-equivalent
energy-complexification patterns or, if you wish, of the
non-equivalent scenarios of a ``quantum catastrophe". Moreover, we
found that once we fix a real $J-$plet of optional parameters
$G_n$ in ansatz~(\ref{lobkov}), the range of the remaining free
parameter $t\in (-\infty,\infty)$ splits in the four specific
subintervals:
 \begin{enumerate}
 \item
in an ``unobservable" regime, some of the energies are not real
(so that $H^{(N)}$ itself is not quasi-Hermitian, QH) at  $t\in
\left (-\infty,t_{(QH)}(G_1, G_2,\ldots,G_J)\right )$;
 \item
the genuine quasi-Hermitian {\em and} ${\cal PT}-$symmetric regime
is encountered in the range of $t \in \left (t_{(QH)}(G_1,
G_2,\ldots,G_J),t_{(PH)}(G_1, G_2,\ldots,G_J)\right )$ where one
stays safely inside ${\cal D}^{(N)}$. The value of the
parity-pseudo-Hermiticity boundary $t_{(PH)}$ is given by the
decision \cite{I} that the matrix $H^{(N)}$ remains real, $\max_n
\xi_n(t) \leq 1$;
 \item
in the next interval of $t \in \left (t_{(PH)}(G_1,
G_2,\ldots,G_J),t_{(H)}(G_1, G_2,\ldots,G_J)\right )$, the matrix
$H^{(N)}$ ceases to be real {\em and} its ${\cal P}-$ (i.e.,
parity-) pseudo-Hermiticity ``strengthens" to an
$\eta-$pseudo-Hermiticity. An $N=4$ illustrative example of the
``modified parity" $\eta$ (which may further vary with $t$) was
given in ref.~\cite{condit};
 \item
Hermitian regime enters the scene at $t \in \left (t_{(H)}(G_1,
G_2,\ldots,G_J),\infty \right )$. One has $\xi_{n}(t)> 1$ at all
$n$ so that all the couplings $g_{n}$ become purely imaginary.

\end{enumerate}

 \noindent
In this setting, one notes a certain complementarity between the
results of refs.~\cite{I} and \cite{condit}. In the former (and
older) text we intended to stay safely inside {\em the closure} of
the domain ${\cal D}^{(N)}$. In this framework we showed that the
choice of the {optimal} proportionality coefficients (\ref{optim})
in eq.~(\ref{lamb}) implies that we can minimize $t_{(QH)}=0$. In
this sense we found the {\em maximal} interaction strengths which
are still compatible with the reality of the spectrum. In
contrast, the scope of paper \cite{condit} was broader and covered
(i.e., sampled, numerically) {\em all} the four eligible intervals
of the auxiliary parameter $t$.

In what follows we shall show that and how the successful
localization of the exceptional boundary point with $t_{(QH)}=0$
in \cite{I} can be extended to a certain approximative
closed-formula description of all the $(J-1)-$dimensional boundary
set $\p {\cal D}^{(N)}$ in a vicinity of this point. This will
definitely clarify, {\it inter alii}, that and how the necessary
physical stability of our model can be guaranteed via a
constructive, leading-order specification of an {\em interior,
strong-coupling} part of the domain ${\cal D}^{(N)}$.

\section{\label{parag3} The method of explicit constructions}

\subsection{\label{par2.3}Guiding example: $N=4$
 \label{idea}  }

In the first nontrivial $N=4$ example we  set $
g_1=\sqrt{3\,(1-\beta)}$, $ g_2=\sqrt{4\,(1-\alpha)}$ and consider
 \ben
 H^{(4)}=
 \left (\begin {array}{cccc}
  -3&\sqrt {3-3\,\beta}&0&0\\
  -\sqrt {3-3\,\beta}&-1&
 2\,\sqrt {1-\alpha}&0\\
 0&-2\,\sqrt {1-\alpha}&1&\sqrt {3-3\,\beta}\\
 0&0&-\sqrt {3-3\,\beta}&3
 \end {array}
 \right )\,
 \een
with $\alpha\,,\beta \in (0,1)$. This leads to the secular
equation
 \be
 {s}^{2}-\left (6\,\beta+4\,\alpha\right )s
 -36\,{\beta}+36\,\alpha+9\,{\beta}^{ 2}=0
 \label{quartea}
 \ee
with the doublet of available elementary roots,
  \be
 s_\pm=3\,\beta+2\,\alpha \pm 2\,\sqrt {3\,\beta\,
 \alpha+{\alpha}^{2}+9\,\beta-9\,\alpha}\,.
 \label{ruty}
  \ee
As long as the energies are square roots of these roots we must
guarantee that $s_\pm \geq 0$. {\it Vice versa}, the latter two
inequalities may be understood as an implicit definition of the
domain ${\cal D}^{(4)}={\cal D}^{(4)}(\alpha,\beta)$. More details
may be found elsewhere \cite{kde}.

In an attempt to clarify the origin of ansatz (\ref{lobkov}) let
us now introduce an auxiliary, redundant parameter $t$ and set
$\alpha=t\,a$ and $\beta=t\,b$, treating $t$ as a radius of the
EEP vicinity and preserving just the leading-order terms in $t$.
Then, conditions $s_\pm \geq 0$ degenerate to the two elementary
rules
 \ben
 b\geq a +{\cal O}(t), \ \ \ \ \ \ \
  a\geq b +{\cal O}(t)\,
  \een
which may be interpreted as a requirement of a fine-tuned balance
between $a$ and $b$ (or $\alpha$ and $\beta$) near the EEP
extreme. We may conclude that our ansatz (\ref{lobkov}) is optimal
and that {\em inside} ${\cal D}^{(4)}$, the value of $\alpha$ can
only differ from $\beta$ in the {\em next} order in the small $t$,
 \be
 \beta=t+B\,t^2\,,\ \ \ \ \ \ \alpha=t+A\,t^2\,.
 \label{ans4}
 \ee
Tractable as an exact change of variables $(\alpha,\beta)
\longrightarrow (A,B)$, such a rule {\em strictly} replaces the
inequalities $s_\pm \geq 0$ by the $t-$parametrized pair of the
necessary and sufficient conditions of the physical acceptability
of $H^{(4)}$,
 \be
 {t}^{2}{{\it {A}}}^{2}+5\,t{\it {A}}
 +3\,{t}^{2}{\it {A}}\,{\it {B}}+4+3\,t
 {\it {B}}+9\,{\it {B}}-9\,{\it {A}}\geq 0\,,
 \label{vica}
 \ee
 \be
 {t}^{2}+2\,{t}^{3}{\it {B}}+{t}^{4}{{\it {B}}}^{2}
 -4\,{t}^{2}{\it {B}}+4
 \,{t}^{2}{\it {A}}\geq 0\,.
 \label{vicbe}
 \ee
In the vicinity of the EEP extreme we can omit the higher-order
terms from eqs.~(\ref{vica}) and (\ref{vicbe}). This reduces this
pair of inequalities to the compact leading-order estimate
 \be
 \frac{4}{9}\ \geq\ A-B\ \geq \, -\frac{1}{4}\,.
 \label{redus}
 \ee
This leading-order rule {defines} ${\cal D}^{(4)}$ {\em reliably}
near the EEP vertex. Indeed, in a graph-drawing scenario we may
suppress the redundancy and fix $t \equiv \beta$ (i.e., set
$B=0$). This converts (\ref{redus}) into the explicit
leading-order formulae which define the two branches of the
boundary $\p {\cal D}^{(4)}$,
 \be
 \alpha^{(upper)}(\beta)=\beta+\frac{4}{9}\,\beta^2+{\cal O}(\beta^3)\,,
 \ \ \ \ \
 \alpha^{(lower)}(\beta)=\beta-\frac{1}{4}\,\beta^2+{\cal
 O}(\beta^3)\,.
 \label{labba}
 \ee
These two curves osculate at EEP so that the domain ${\cal
D}^{(4)}$ appears sharply spiked near this extreme.

It makes sense to believe that such a geometric property of the
domain ${\cal D}$ is generic. At the higher matrix dimensions $N$,
precisely this hypothesis has been formalized by the tentative
ansatz~(\ref{lobkov}). For a graphical illustration of its
consequences we decided to sample the $N=4$ spectrum in Figure~1.
At $B=0$ it compares the quadruplet of the energies which are real
at all $t \geq 0$ (and which correspond to the ``admissible"
choice of the rescaled coupling $A=4/9-20/100$ which lies safely
inside ${\cal D}^{(4)}$) with another quadruplet obtained at a
``forbidden" $A=4/9+2/100$ [i.e., slightly outside ${\cal
D}^{(4)}$, violating the upper bound in eq.~(\ref{redus})] which
remains all complex at all the positive small $t<t_{(QH)}(A)$.

%
%
%
%
%

\begin{figure}[t]                     
\begin{center}                         
\epsfig{file=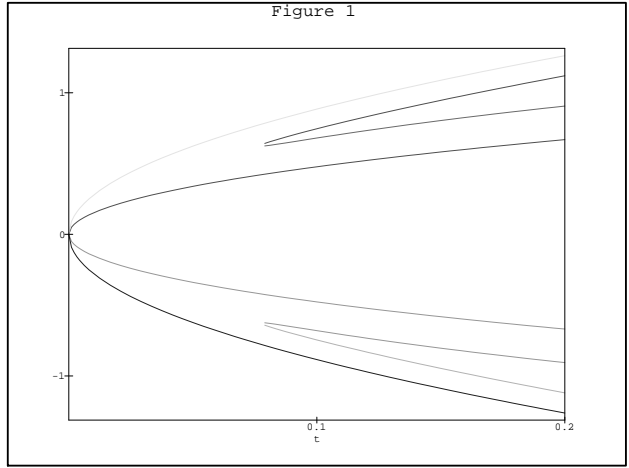,angle=270,width=0.6\textwidth}
\end{center}                         
\vspace{-2mm} \caption{The $t-$dependence of energies, with two
choices of $A$ at $N=4$.
 \label{obr2a}}
\end{figure}

%

\subsection{\label{par2.5} Inessential changes at $N=5$}


At the first nontrivial odd dimension $N=5$ with $g_1=b=2\,\sqrt
{1-\beta}$, $\beta \in (0,1)$ and $g_2=a=\sqrt {6\,(1-\alpha)}$,
$\alpha \in (0,1)$, our model
 \ben
 H^{(5)}=
 \left (\begin {array}{ccccc}
  -4& 2\,\sqrt {1-\beta}&0&0&0\\
  - 2\,\sqrt {1-\beta}&-2&\sqrt {6-6\,\alpha}&0&0\\
 0&
  -\sqrt {6-6\,\alpha}&0&\sqrt {6-6\,\alpha}&0\\
  0&0& -\sqrt {6-6\,\alpha}&2&2\,\sqrt {1-\beta}\\
 0&0&0&- 2\,\sqrt {1-\beta}&4
 \end {array}
 \right )\,
 \een
leads to the secular polynomial of the fifth degree in $E$ which
is divisible by $E$. Thus, one of the roots, viz., the energy
$E_2=0$ may be treated as trivial. The remaining four energies $E$
may be computed from the two roots $s=E^2$ of the quadratic
equation
 \be
 s^2-P_1^{(5)}(g_1, g_2)\,s^{}
 +P_2^{(5)}(g_1,  g_2)=0
 \label{fivea}
  \ee
where one easily evaluates
 \ben
 P_1^{(5)}(g_1, g_2)=
 8\,\beta+12\,\alpha\,,
 \ \ \ \ \ \ \
  P_2^{(5)}(g_1,  g_2)=
 48\,\alpha\,\beta-144\,
 \beta+144\,\alpha+16\,{\beta}^{2}\,.
 \een
We may skip the details -- near EEP the construction of ${\cal
D}^{(5)}$ is strikingly analogous to its $N=4$ predecessor.

\subsection{\label{parag3.3} Inessential changes at $N=6$}


At $N=6$, three parameters $\alpha,\beta,\gamma \in (0,1)$ enter
the three coupling constants
 \ben
 g_1=c=\sqrt{5\,(1-\gamma)}\,,\ \ \ g_2=b= 2\,\sqrt{2\,(1-\beta)}\,,
 \ \ \ g_3=a= 3\,\sqrt{1-\alpha}
 \label{sol3}
 \een
which specify the dynamics via the Hamiltonian
  \be
 H^{(6)}
 =\left [\begin {array}{cccccc}
  -5&g_1&0&0&0&0\\
 -g_1& -3&g_{2}&0&0&0\\
 0&-g_{2}&-1&g_3&0&0
 \\
 0&0&-g_{3}&1&g_{2}&0
 \\
 0&0&0&-g_2&3&g_{1}\\
 0&0&0&0&-g_{1}&5
 \end {array}\right ]\,.
 \label{hamm}
 \label{6IPTS}
 \ee
The shape of the associated domain of  quasi-Hermiticity ${\cal
D}={\cal D}(a,b,c) $ can be deduced from the secular equation
 \be
 \det\left ( H^{(6)}-E\,I^{(6)}\right ) =
 s^3-3\,P_1^{(6)}\,s^2+3\,P_2^{(6)}\,s-P_3^{(6)}=0\,,\ \ \ \ \ \ \
 s=E^2\,
 \label{naskuba}
 \ee
(notice the slightly modified notation) re-written as the relation
 \ben
 \left [s-s_1(a,b,c)\right ]\,
 \left [s-s_2(a,b,c)\right ]\,
 \left [s-s_3(a,b,c)\right ]=0\,
 \een
between its roots $s_k(a,b,c)$. As long as they define the
energies $E_{\pm k}=\pm \sqrt{s_k(a,b,c)}$, all of them must be
nonnegative in  ${\cal D}^{(6)}$. For this reason, we have to
satisfy three requirements $P_k^{(6)}\geq 0$, $k=1,2,3$ plus a
certain slightly more complicated fourth condition (with the
derivation left to the reader as an exercise).

From a geometric point of view, the construction of the boundary
$\p {\cal D}^{(6)}$ of the physical domain remains similar to its
$N=4$ predecessor. Thus, equation
 \ben
 P_1^{(6)}=-\left ({a}^{2}+2\,{b}^{2}+2\,{c}^{2}-35\right )/3=0
 \een
determines the ellipsoid which circumscribes the domain ${\cal
D}^{(6)}$. In contrast, the geometric interpretation of the
further circumscribed surfaces (given by equations $P_2^{(6)}=0$
etc) is much less straightforward. For this reason we intend to
employ ansatz (\ref{lobkov}) and to replace the variables
$\alpha,\beta,\gamma$ by the new triplet $A,B,C$ defined by the
$J=3$ version of the recipe,
 \be
  \alpha = t+t^2+A\,t^3 \,,
 \ \ \ \ \
  \beta = t+t^2+B\,t^3 \,,
 \ \ \ \ \
  \gamma = t+t^2+C\,t^3 \,.
 \label{ans6}
  \label{malik}
 \ee
The auxiliary variable $t$ is redundant but useful because we
intend to keep it small.

\section{\label{parag4} The method of rescaled couplings}

Our above exact result (\ref{ans4}) can be reinterpreted as a
tentative ansatz with the obvious generalization (\ref{lobkov}).
The use of the latter rule requires an assumption of the smallness
of $|t|\ll 1$. This could simplify our secular
eq.~(\ref{seculare}),  near the EEP extremes at least.

\subsection{Guiding example: $N=4$ \label{par2.4} }

In the new notation our secular equation (i.e., our implicit
definition of the $N=4$ energies) degenerates to the leading-order
relation
 \ben
 s^2-10\,t\,s
 +\left (36\,A-36\,B+9\right ){t}^{2}+{\cal O}(t^3)=0\,.
 \een
After we introduce a new coupling parameter
$\omega=\omega^{(J)}=36\,(A-B)$, this equation for the unknown
quantity $L=s/t$ acquires a transparent $t-$independent
leading-order form which may be partially factorized,
 \be
 L^2-10\,L
 +9 + \omega=
 (L-1)\,(L-9) + \omega=0\,.
 \label{bette}
 \ee
It is important that the roots of eq.~(\ref{bette}) are known
exactly,
 \ben
 L_\pm = 5\pm \sqrt{16-\omega}\,.
 \een

\subsubsection{The leading-order localization of the boundary
$\p {\cal D}^{(4)}$}

Obviously, the growth of $\omega>0$ beyond its ``upper limit"
$\omega_{UL}=16$ makes all the four roots $L$ (i.e., all the
related leading-order energies) complex. In the alternative
scenario, the decrease of $\omega<0$ below its ``lower limit"
$\omega_{LL}=-9$ makes just one of the roots (viz., $L_-$)
negative.

Both these estimates coincide with the above-derived formula
(\ref{redus}). We may conclude that the use of our
perturbation-type ansatz (\ref{ans4}) reproduces, completely, the
leading-order information about the spiked shape of the boundary
$\p {\cal D}^{(4)}$ near the EEP extreme.

We could also speak about the quadruplet of the $t-$dependent
energies $E_0(t)$, $\ldots$, $E_3(t)$ studied in the
strong-coupling regime. This dynamical regime is characterized by
the small auxiliary  quantities $t$ which, in effect, measure the
``distance" of our model $H^{(4)}$ from its EEP $t=0$ extreme. In
such an alternative language, the pair of $E_1(t)$ and $E_2(t)$
will complexify somewhere near $\omega \sim \omega_{LL}$, etc (cf.
\cite{condit} for a more complete discussion and for a
combinatorial classification of all the possible scenarios of
complexification at any~$N$).

\subsubsection{A linearized localization of
$\p {\cal D}^{(4)}$ at small $\omega$ }

In the symmetric-coupling regime with $A=B$ one stays safely
inside the ``physical" interior of ${\cal D}^{(4)}$. Once we
assume that the difference $\omega \sim A-B$ is small, the
extraction of all the energy roots becomes facilitated. In the
leading-order approximation their $\omega-$dependence remains
linear,
 \be
 L_-=1+\frac{\omega}{8}+{\cal O}(\omega^2)\,,
 \ \ \ \ \ \ \ \
 L_+=9-\frac{\omega}{8}+{\cal O}(\omega^2)\,.
 \label{appp}
 \ee
These formulae still offer a {\em qualitatively correct}
perturbative explanation of the {\em  complexification pattern} of
the energies. Indeed, even from the oversimplified approximation
$s=E^2=L\,t+{\cal O}(t^2)$ using closed formulae~(\ref{appp}) one
can deduce the rough estimates of the critical
$\omega_{UL}^{(1)}=32$ (or cca $23.4$ in the second-order
approximation in $\omega$) and $\omega_{LL}^{(1)}=-8$ (or $-9.37$
in the second-order approximation), yielding even a reasonably
good quantitative prediction.

\subsubsection{A systematic build-up of higher-order corrections}

Whenever we keep the $t-$dependent version of our $N=4$ secular
eq.~(\ref{quartea}) in full precision, we can construct the
perturbation series for the energies at the small $t$ in the
standard manner, with
 \ben
 E_3 = -E_0=\sum_{k=0}^\infty\,t^{k+1/2}\,E_3^{(k)}\,,\ \ \ \ \ \
 E_2 = -E_1=\sum_{p=0}^\infty\,t^{p+1/2}\,E_2^{(p)}\,
 \een
and with
 \ben
 E_3^{(0)}=
 \sqrt {2\,\sqrt {9\,B-9\,A+4}+5}\,\ ,
 \een
 \ben
  E_3^{(1)}={\frac {1}{2\,\sqrt {2\,\sqrt {9\,B
 -9\,A+4}+5}}}
 \,
 \left({\frac {3\,B+5\,A}{\sqrt
 {9\,B-9\,A+4}}}+3\,B+2\,A\right)\,\ ,
 \een
etc., or with
 \ben
 E_2^{(0)}=
 \sqrt {5-2\,\sqrt {9\,B-9\,A+4}}\,\ ,
 \een
 \ben
  E_2^{(1)}={\frac {1}{2\,\sqrt {5-2\,\sqrt {9\,B-9
\,A+4}}}}\,\left(2\,A+3\,B-{\frac {3\,B+5\,A}{\sqrt
{9\,B-9\,A+4}}}\right)\,\ ,
 \een
etc. These formulae offer another source of insight in the
mechanisms of complexification of the spectrum at the boundary $\p
{\cal D}^{(4)}$.

\subsection{\label{par5.1}A change of the pattern at $N=5$
 }

The insertion of  ansatz (\ref{ans4}) in  secular equation
(\ref{fivea}) gives
 \ben
 P_1^{(5)}(b,a) = 20\,t+8\,{t}^{2}B+12\,{t}^{2}A \,.
  \een
This confirms our expectations that the necessary non-negativity
of this expression is guaranteed at all the not too large $|A|,
|B| \ll 1/t$. Similarly we are able to evaluate and demand that,
in full precision,
 \ben
 P_2^{(5)}(b,a) =144\,{t}^{2}A-144\,{t}^{2}B
 +64\,{t}^{2}+16\,{t}^{4}{B}^{2}
 +48\,{t}^{3}A+48\,{t}^{4}A\,B+80 \,{t}^{3}B\geq 0\,.
 \een
Obviously, the higher-order corrections may only be needed quite
far from the EEP extreme. In the dominant order in $t$ we may
conclude  that
 \ben
 s^2-20\,t\,s
 +\left (144\,A-144\,B+64\right ){t}^{2}+{\cal O}(t^3)=0\,.
 \een
Once we put $s=L\,t+{\cal O}(t^2)$ again, we get
 \be
 L^2-20\,L
 +64 + \varepsilon=(L-4)\,(L-16) + \varepsilon=0\,,\ \ \ \ \ \ \ \
 \varepsilon=144\,
 \left (A-B\right )\,.
 \label{bette5}
 \ee
In the regime with a small $\varepsilon$ (of any sign), we easily
get the leading-order $\varepsilon-$dependence of both the roots,
 \ben
 L_-=4+\frac{\varepsilon}{12}+{\cal O}(\varepsilon^2)\,,
 \ \ \ \ \ \ \ \
 L_+=16-\frac{\varepsilon}{12}+{\cal O}(\varepsilon^2)\,.
 \een
Marginally, it is amusing to notice that in the trivial case with
$A-B=\varepsilon=0$ the positive doublet of energies (as well as
its negative counterpart) has a tendency of moving directly to the
corresponding weak-coupling harmonic-oscillator limit. Our
leading-order approximation $E_\pm = +\sqrt{(10\pm 6)\,{t}+{\cal
O}(t^2)}$ behaves as if being, paradoxically, exact at $t=1$.

\subsubsection{Towards the boundary $\p {\cal D}^{(5)}$}

In a way paralleling our above $N=4$ considerations let us switch
to a larger, positive $\varepsilon \sim A-B>0$. The two dominant
energies $E_3\geq E_2>0$ will then decrease and increase with the
growth of $\varepsilon$, respectively. One can predict an ultimate
coincidence and a subsequent complexification of this doublet at a
non-vanishing Kato's exceptional point localized at some
``boundary coordinate" $t=t_{(QH)}>0$ of quasi-Hermiticity.

For the opposite, negative and decreasing $\varepsilon<0$ the
approximative results are similar and do not lead to any
contradictions. We may summarize that near the shared EEP maximum
of $g_1$ and $g_2$, the growth of the difference between $\alpha$
and $\beta$ would lead us out of the physical domain ${\cal
D}^{(5)}$. In the other words, the boundary $\p {\cal D}^{(5)}$
near EEP is of a sharply spiked form as well.

\section{Extrapolation hypothesis \label{par}}

Let us now turn attention to the models with $J=3$. We intend to
offer some quantitative arguments in favor of the intuitive idea
that the exceptional-point boundaries $\p {\cal D}^{(6)}$ and  $\p
{\cal D}^{(7)}$ have a form of a surface of a deformed cube with
protruded, razor-sharp edges (= double exceptional points) and
with the spiked, strong-coupling vertices (i.e., triple
exceptional points).

We expect that such an intuitively transparent geometric
interpretation of the shape of ${\cal D}^{(6)}$ and ${\cal
D}^{(7)}$ will provide a firm ground for extrapolations towards
higher dimensions in subsequent sections of this paper.

\subsection{\label{par4.2}Expectable observations at $N=6$ }

After transition from $N=4$ to $N=6$, symbolic manipulations on a
computer become rather lengthy. Still, they enable us to evaluate,
in closed form, the approximate secular-equation coefficients
 \ben
 P_1^{(6)}=P_1^{(6)}(c,b,a)= 35\,t+35\,{t}^{2}
 +{\cal O}(t^3)\,,
  \een
 \ben
 P_2^{(6)}=P_2^{(6)}(c,b,a)=259\,{t}^{2}
 +\left (216\,{\it {A}}+518+144\,{\it {B}}-360\,{\it {C}}
 \right ){t}^{3}
 +{\cal O}(t^4)\,
  \een
and
 \ben
 P_3^{(6)}=P_3^{(6)}(c,b,a)=\left (3600\,{\it {A}}
 +1200\,{\it {C}}-4800\,{\it {B}}+225\right ){t}^{3}
 +
 \ \ \ \ \ \ \ \
 \ \ \ \ \ \ \ \
 \ \ \ \ \ \ \ \
 \een
 \ben
 \ \ \ \ \ \ \ \ \ \ \ +
 \left (3600\,{\it {B}}-1800\,{\it {C}}
 +675-1800\,{\it {A}}\right )
  {t}^{4}
 +{\cal O}(t^5)\,.
 \een
The resulting shortened leading-order secular equation is finally
obtained and reads
 \ben
 s^3-35\,t\,s^2
 +259\,{t}^{2}\,s
 -\left (225
 -\omega^{(3)}\right ){t}^{3}
 =0\,,
 \ \ \ \ \ \ \ \
 \omega^{(3)}=1200\,
 (-3\,A+4\,B-C)
 \,.
 \een
After we put $s=E^2=L\,t+{\cal O}(t^2)$, its partial factorization
remains feasible,
 \be
 (L-1)\,(L-9)\,(L-25) + \omega=0\,.
 \label{bette6}
 \ee
In a way paralleling the previous $N=4$ example, a transparency of
this equation facilitates the analysis of the relationship between
the variations of the coupling-dependent quantity $\omega \sim
4\,B-C-3\,A$ and of the physical spectrum of energies
$E_0=-E_5\leq E_1=-E_4 \leq E_2=-E_3$, inside the domain of its
reality at least.

\subsubsection{Towards the boundary $\p {\cal D}^{(6)}$}

The discussion is easier at the small $\omega^{(3)}$ where all the
three roots of eq.~(\ref{bette6}) are almost linear in $\omega$,
 \ben
 L_1=1-\frac{\omega}{8\cdot 24}+ {\cal O}(\omega^2)\,,
 \ \ \ \ \
 L_2=9+\frac{\omega}{8\cdot 16}+ {\cal O}(\omega^2)\,,
 \ \ \ \ \
 L_3=25-\frac{\omega}{16\cdot 24}+ {\cal O}(\omega^2)\,.
 \een
At $\omega\approx 0$ and at the smallest $t\neq 0$ we have
$E_0\sim -5\,\sqrt{t}$, $E_1\sim -3\,\sqrt{t}$,  $E_2\sim
-\sqrt{t}$, $E_3\sim \sqrt{t}$, $E_4\sim 3\,\sqrt{t}$ and $E_5\sim
5\,\sqrt{t}$. With the decrease of $\omega<0$ the levels $E_1$ and
$E_2$ (and also $E_3$ and $E_4$) get closer to each other and, in
a way which parallels the similar observation made at $N=4$, they
finally complexify at the critical $\omega_{(LL)}\approx
-323.1387184$ near $E_3 \approx E_4\approx 2.147400716\,\sqrt{t}$.
In the light of the obvious fact that the growth of every coupling
$g_k$ causes an attraction of the corresponding levels
\cite{PLB2}, the interpretation of the above rule is easy and
consistent because the decrease of $\omega$ means not only a
smaller $B<(3\,A+C)/4$ but also, at any given $t$, $g_1$ and
$g_3$, an enhancement of the coupling $g_2$ between the levels
which complexified.

In the opposite direction, the increase of $\omega>0$ causes the
decrease of the smallest root $L_s(\omega)\ (=E^2_{2,3}/t)$ of
eq.~(\ref{bette6}) while the other two roots  $L_m(\omega)\
(=E^2_{1,4}/t)$ and $L_d(\omega)\ (=E^2_{0,5}/t)$ start
approaching each other. In this way, the first and dominating
complexification involves the pair $E_{2,3}$ (with $E_{2,3}\approx
0$) because it takes place exactly at $\omega_{(UL)}=225$, i.e.,
much earlier than the subsequent ultimate completion of all the
complexification process with $E_5 \approx E_4\approx
4.326893054\,\sqrt{t}$ near $\omega_{(UL[4,5])}\approx
1081.657240$.

\subsection{\label{par5.2}Expectable changes at $N=7$ }

With $N=7$ and with the $J=3$ rule (\ref{ans6}), we employ the
symbolic manipulations on a computer and get five terms in
 \ben
 P_1^{(7)}(a,b,c)=56\,t+56\,{t}^{2}+{\cal O}(t^3)
  \een
sixteen terms in
 \ben
 P_2^{(7)}(a,b,c)=784\,{t}^{2}+{\cal O}(t^3)
  \een
and 33 terms in
 \ben
 P_3^{(7)}(a,b,c)=\left (7200\,C-21600\,B
 +14400\,A+2304\right )
{t}^{3}+{\cal O}(t^4)\,.
 \een
The resulting shortened leading-order secular equation reads
 \ben
 s^3-56\,t\,s^2
 +784\,{t}^{2}\,s
 -\left (2304
 -\varepsilon\right ){t}^{3}
 =0\,,
 \ \ \ \ \ \ \ \
 \varepsilon=7200\,
 (-C+3\,B-2\,A)
 \,.
 \een
With $s=L\,t$ it factorizes as follows,
 \be
 (L-4)\,(L-16)\,(L-36) + \varepsilon=0\,.
 \label{bette7}
 \ee

\subsubsection{Towards the boundary $\p {\cal D}^{(7)}$}

The three closed solutions of eq.~(\ref{bette7}) may be expanded
in the powers of $\varepsilon$,
 \ben
 L_1=4-\frac{\varepsilon}{384}+ {\cal O}(\varepsilon^2)\,,
 \ \ \ \ \
 L_2=16+\frac{\varepsilon}{240}+ {\cal O}(\varepsilon^2)\,,
 \ \ \ \ \
 L_3=36-\frac{\varepsilon}{640}+ {\cal O}(\varepsilon^2)\,.
 \een
A balanced return to the unperturbed values occurs now along a
``middle line" with $\varepsilon =0$, i.e., for $B=(2\,A+C)/3$.

For the diminished $B$s (i.e., for a stronger coupling $b=g_2$
between $E_1$ and $E_2$), we are getting closer to the boundary
surface along which the levels $E_1$ and $E_2$ merge and
subsequently complexify. This behaviour is confirmed by our
formula because the shift $\varepsilon$ becomes negative in such a
scenario.

The growth of $\varepsilon > 0$, on the contrary, may be assigned
to the diminished $A$ or $C$. In the former case the growth of
$a=g_1$ implies that $E_1 \to 0$ while the alternative growth of
$c=g_3$ results, naturally and expectedly, in the merger of $E_2$
with $E_3$.

\section{\label{par4}Arbitrary even dimension and the localization of
boundaries $\p {\cal D}^{(2J)}$ in the EEP vicinity \label{par3}}
%
%

We saw that in many a respect one should pay separate attention to
the models with even and odd~$N$. Thus, we shall search for a
successful extrapolation of our low$-N$ ``experiments" in two
separate sections, starting with the family where $N=2J$.

Our first assumption concerns the values of the rescaled couplings
$G_n$ (abbreviated, occasionally, as $A, B$ etc) which will be
assumed bounded while the scaling parameter $t$ itself will be
assumed small.

Our second, main assumption is that our older eqs.~(\ref{bette})
and eq.~(\ref{bette6}) are just the respective $N=4$ and at $N=6$
special cases of the general leading-order secular equation
 \be
 \left (L-1^2\right )\,
 \left (L-3^2\right )\,
 \left (L-5^2\right )\,\ldots\;
 \left (L-[2J-1]^2\right )\,
 + \omega^{(J)}=0\,.
 \label{betteJe}
 \ee
We tested this hypothesis together with the ansatz (\ref{lobkov})
by extensive symbolic manipulations which confirmed its validity
at {\em all} the integers $N=2J$ in a way documented in Table~1.
This Table summarizes also the corresponding resulting explicit
formulae for the $J-$dependent quantities $\omega^{(J)}$. Using
the same extrapolation philosophy as above, we may extract the
universal algebraic definition of the coupling-characterizing
expression $\omega^{(J)}$ as given by the following extrapolated
conjecture {\em at all the positive integers $J$},
 \be
 \omega^{(J)}
 = 2\,(2J-1)\,(2J-1)!\,
 \sum_{n=1}^J\,(-1)^{J-n+1}\,
 \left (
 \ba
 2\,J-2\\
 J-2+n
 \ea
 \right )\,
 \theta(n)
 \,G_n\,.
 \label{formu}
 \ee
This is our first main result deduced via an interpolation of the
computed coefficients at the first few dimensions $N$,
complemented by a subsequent (and, of course, much easier)
verification of the hypothesis at a number of the higher $N$.
Formula (\ref{formu}) contains just the usual combinatorial
numbers (given by the standard Pascal-triangle recurrences) and
the anomalous scaling factor $\theta(n)$ which is equal to one at
$n>1$ and to one half at $n=1$.

The geometric interpretation of formula (\ref{formu}) could be
discussed in an immediate parallel with the texts on the special
cases with $N=4$ and $N=6$ in the respective paragraphs
\ref{par2.4} and \ref{par4.2}. At the general $J$, a careful
inspection of the first few secular eqs.~(\ref{betteJe}) seems to
indicate that at the smallest $t$s, the complexification of the
spectrum {\em always} proceeds

\begin{itemize}

 \item
through the single merger, at the zero energy,  of the ``middle"
levels $E_{J-1}$ and  $E_{J}$ (at some $\omega_{(UL)}^{(J)}>0$ for
$J=1, 3, 5, \ldots$ and at some $\omega_{(LL)}^{(J)}<0$ for
$J=2,4,6, \ldots$),

 \item
through the two simultaneous, symmetric mergers of the negative
$E_{J-3}$ with $E_{J-2}$ and of the positive $E_{J+1}$ with
$E_{J+2}$ (at $\omega_{(LL)}^{(J)}<0$ for $J=1, 3, 5, \ldots$ and
at $\omega_{(UL)}^{(J)}>0$ for $J=2,4,6, \ldots$).

\end{itemize}

\section{\label{par5}All odd dimensions and the
boundaries $\p {\cal D}^{(2J+1)}$ in the EEP vicinity
 }


When the dimension is odd, $N=2J+1$,  the application of the
ansatz (\ref{lobkov}) simplifies the problem significantly as
well. This ansatz can be perceived again as an equivalence
transformation based on a mere change of the variables, $g_n \to
G_n$. It maps the domain ${\cal D}^{(N)}(g_1, \ldots,g_J)$ of
quasi-Hermiticity of $H=H^{(2J+1)}(g_1, \ldots,g_J)$ into another,
equivalent manifold ${\cal D}^{(2J+1)}(G_1, \ldots,G_J)$  of the
acceptable parameters in $H=H^{(2J+1)}(G_1, \ldots,G_J)$.

For all the odd dimensions $N$ we may ignore the persistent
``middle" bound-state energy $E^{(2J+1)}_{J}=0$. {\it Mutatis
mutandis}, this enables us to use the same $J$ and to apply the
same (or at least very similar) geometric, algebraic and analytic
considerations as above.

One of the most visible differences between the models $H^{(2J)}$
and  $H^{(2J+1)}$  may be traced to the fact that in place of the
(hyper)ellipsoids pertaining to the even dimensions we have to
deal with the {\em simpler} (hyper)spheres ${\cal S}^{(N)}$ at all
$N=2J+1$. In the same comparison, in contrast, the leading-order
secular equation
 \be
 \left (L-2^2\right )\,
 \left (L-4^2\right )\,
 \left (L-6^2\right )\,\ldots\;
 \left (L-[2J]^2\right )\,
 + \varepsilon^{(J)}=0\,
 \label{betteKa}
 \ee
pertaining to the odd $N=2J+1$ is slightly more complicated. This
is demonstrated by Table~2 showing that the general formula for
the factors $\varepsilon^{(J)}$
 \be
 \varepsilon^{(J)}
 = 2\,(2J-1)\,(2J)!\,
 \sum_{n=1}^J\,(-1)^{J-n}\,C_{(n)}^{(J)}\,G_n\,
 \label{definit}
 \ee
is perceivably less explicit. The reason is that the elementary
combinatorial coefficients in eq.~(\ref{formu}) are replaced now
by no-name integer coefficients $C_{(n)}^{(J)}$.

Fortunately, the latter coefficients can be understood as an
immediate generalization of the current combinatorial numbers
since, firstly, they exhibit a left-right antisymmetry
$C_{(1-n)}^{(J)}=-C_{(n)}^{(J)}$ at $n=1,2,\ldots$ and, secondly,
these coefficients may be generated by the recurrences
 \be
 C^{(J)}_{(n)}=
 C^{(J-1)}_{(n-1)}+
 2\,C^{(J-1)}_{(n)}+
 C^{(J-1)}_{(n+1)}\,
 \label{pascal}
 \ee
from the initial row $C_{(n)}^{(J)}$ at $J=1$ which contains a
unit at $n=1$ and zero for $n>1$ (cf. Table 3). Thirdly, a
decomposition of the sum (\ref{pascal}) in the two steps specifies
the coefficients $C_{(n)}^{(J)}$ as a subset of the triangle
generated by the usual combinatorial three term recurrences. The
latter generation pattern is sampled in Table~4. One notices that
both the three-term recurrences and their illustrative Table~4
differ from their standard Pascal-triangle predecessors merely in
an anomalous, left-right antisymmetric initialization.

Once we return to our leading-order energies $E_n^{(2J+1)}$ we may
say, on the basis of eq.~(\ref{betteKa}), that their {\em
leading-order} complexification {\em always} seems to proceed in
one of the following two ways:

\begin{itemize}

 \item
through the two simultaneous, symmetric mergers of the negative
$E_{J-2}$ with $E_{J-1}$ and of the positive $E_{J+1}$ with
$E_{J+2}$ (at $\varepsilon_{(LL)}^{(J)}<0$ for $J=1, 3, 5, \ldots$
and at $\varepsilon_{(UL)}^{(J)}>0$ for $J=2,4,6, \ldots$).

 \item
through the two simultaneous, symmetric mergers of the negative
$E_{J-3}$ with $E_{J-2}$ and of the positive $E_{J+2}$ with
$E_{J+3}$ (at $\varepsilon_{(UL)}^{(J)}>0$ for $J=1, 3, 5, \ldots$
and at $\varepsilon_{(LL)}^{(J)}<0$ for $J=2,4,6, \ldots$).

\end{itemize}

 \noindent
In the other words, the implicit definition (\ref{betteKa}) of the
spectrum has a similar geometric interpretation and implications
for the shape of the boundary $\p {\cal D}^{(2J+1)}$ as its
even-dimensional predecessor (\ref{betteJe}) did.

\section{\label{parag8} Discussion}

\subsection{The real Kato's exceptional points}

In the Kato's book \cite{Kato}, certain $\kappa-$dependent
one-parametric families of $N$ by $N$ matrices were discussed,
with elements defined as holomorphic functions of $\kappa$ in a
complex domain $D_0$. It has been noticed there that up to not too
many ``exceptional points" $\kappa^{(EP)}\in D_0$, the number of
eigenvalues of any such matrix $H(\kappa)$ is equal to a constant.

One of the Kato's illustrative two-by-two examples possessing just
a real pair of the Kato's exceptional points was discussed in our
recent paper \cite{PLB2}. In his/our example ($=H^{(2)}$ in our
present notation) we worked just with the real $\kappa\equiv g_1$
and identified the Kato's exceptional points as points of the
{boundary} of the corresponding quasi-Hermiticity domain,
$\kappa^{(EP)} \in \p{\cal D}^{(2)}$.

This was one of the important motivations of our continuing
interest in the structure of the sets $\p {\cal D}^{(N)}$ of the
real exceptional points attached, in particular, to the real chain
models $H^{(N)}$ at the higher matrix dimensions $N$. In these
particular models, the sets of the real Kato's exceptional points
(i.e., the boundary sets $\partial {\cal D}$ of the models in
question) exhibit a nice and transparent hierarchical pattern of
confluence.

Among all the multiple exceptional points, an extreme is
represented by the $J-$tuple confluence of these points at the
strong-coupling extremal EEP vortices \cite{I}. Of course, all the
similar multiple confluence(s) of the exceptional-point
hypersurfaces of various dimensions are in a close correspondence
with some physical critical phenomena. We might emphasize that in
our family $H^{(N)}$, a fine-tuning mechanism emerges which aligns
the values of the physical couplings in a manifestly
non-perturbative though still analytically tractable and
presumably also generic and, in its qualitative aspects, not too
model-dependent manner.

\subsection{Separable anharmonicities}

On the more practical side of our present considerations and
constructions it is worth emphasizing that the Kato's abstract
models of a generic dependence of the eigenvalues on a {\em
single} variable parameter in $H=H(\kappa)\neq H^\dagger$ need not
always offer a better qualitative understanding of the situation.

For an explicit illustrative example we may return to the
differential-equation example of paper~\cite{recenze}. An
inspection of the Table 1 of this paper reveals that once we vary
just an arbitrarily chosen free parameter $\alpha$ (as defined in
eq.~(\ref{aryeh}) above), no obvious pattern is detected in the
complexifications {\em and} decomplexifications of the energy
levels which occur in the Table in an apparently unpredictable and
more or less chaotic manner.

What we would recommend in similar situations would be, in the
light of our present experience, a deeper inspection of the
separate matrix elements themselves, followed by a subsequent
tentative re-definition of the spectrum as depending on some {\em
more} (say, $J$) {\em independently} variable parameters in
$H=H(\kappa_1, \kappa_2, \ldots,\kappa_J)\neq H^\dagger$. Only
then one would have a right to expect that one finds an ``optimal"
linear combination of $\kappa_j$s which wouldn't cross the global
boundary $\p {\cal D}$ at random or, at least, which would stay
{\em locally} more or less perpendiculat to $\p {\cal D}$.

One could expect the existence of an ``optimal" size $J$ of the
set of the auxiliary parameters in many other manifestly
non-Hermitian models based on the use of a complex potential
$V(x)$. Unfortunately, it is difficult to predict how such a type
of analysis would prove efficient in practice. In this sense a
return could be recommended to many older papers on the subject
\cite{Caliceti}.

The problem looks interesting even as a purely mathematical puzzle
of survival of the reality of the energies (i.e., in principle, of
the observability of the system) {\em after} real potentials are
replaced by their complex ${\cal PT}-$symmetric versions. An
interest in such an apparent paradox has been evoked almost ten
years ago \cite{BB,BM} but it took several years before the
necessary {\em rigorous} proofs of the reality of the spectrum
became available \cite{DDT}.

Bad news is that too often, the reality proofs are fairly
complicated \cite{DDTb}. In parallel, good news is that once all
the energies stay real, one returns easily to the standard
principles and formalism of Quantum Mechanics by using an {\it ad
hoc} scalar product in the physical Hilbert space ${\cal H}$. This
means that one merely replaces the usual overlaps $\langle
\phi\,|\,\psi \rangle$ by their generalizations $\langle \phi\,|
\,\Theta\, |\,\psi \rangle $ (this idea belongs to Scholtz et al
\cite{Geyer}).

Although, as a rule, the physical Hilbert-space metric
$\Theta=\Theta^\dagger > 0$ proves complicated and non-local,
people often succeed in its (e.g., perturbative \cite{Batal})
construction. Of course, the explicit specification of the {\em
whole} ``physical" domain ${\cal D}$ seems to be an even more
difficult task. That's why we recommended here the use of the
finite-dimensional prototype matrices $H^{(N)}$.

It is obvious that our knowledge of the boundaries $\p {\cal D}=\p
{\cal D}(H)$ of the physical consistence of a generic ${\cal
PT}-$symmetric Hamiltonian  is necessary for any reliable physical
prediction or for a consistent probabilistic interpretation of the
results of quantum measurements. This boundary may have a fairly
complicated shape even in the simplest, exactly solvable
models~\cite{broken}. At the same time we believe that the present
constructive clarification of the geometric structure of the
corresponding ``prototype physical horizons" $\p {\cal
D}^{(N)}\left (H^{(N)}\right )$ (where even the diagonalizability
of the matrices $H^{(N)}$ themselves gets lost!) could accelerate
the progress in the analysis of the more common $H$s defined via
potentials.

\section{Summary \label{parsum}}

Without any limitations imposed upon $N$, the series of {\em all}
the models $H^{(N)}$ has been shown to admit an innovative
strong-coupling approximative treatment. In particular, the
underlying polynomial secular equations were shown to degenerate
to the closed leading-order form in the strong-coupling dynamical
regime.

An appropriate technical tool for our simplification of the
chain-model bound-state problem in question has been found in {\em
ad hoc} perturbation ansatz (\ref{lobkov}) which reparametrizes
all the coupling constants in a way inspired by their available
closed-form $t\to 0$ limit known from our previous paper \cite{I}.

Also our present methods were inspired by the same reference -- we
further developed the application of the brute-force
symbolic-manipulation techniques as well as of the systematic
formulations and verifications of extrapolation hypotheses.
Marginally, we may add that one of our results, viz., the implicit
leading-order formula for the energies in odd dimensions $N=2J+1$
required a slightly unusual though straightforward and immediate
generalization of the standard binomial coefficients. Our new
combinatorial coefficients were shown closely related to a pair of
alternative generalizations of the popular Pascal triangle.

Via a  deeper analysis of the leading-order secular polynomials we
were, subsequently, able to demonstrate, in the whole
strong-coupling dynamical regime of $H^{(N)}$,  that and how the
determination of the asymptotic strong-coupling parts of the
domains of quasi-Hermiticity ${\cal D}^{(N)}$ degenerates to
elementary approximative formulae at all the dimensions $N$.

Our main conclusion is that our class of models $H^{(N)}$ could be
understood, in many a respect, as representing or mimicking many
generic features of the general ${\cal PT}-$symmetric and
pseudo-Hermitian Hamiltonians. In this sense, in particular,  our
present quantitative description of the mechanism of
complexification of the energies could find an important
application as a guide to our understanding of the change of the
dynamical regime called the spontaneous breakdown of ${\cal
PT}-$symmetry \cite{broken}.

\vspace{5mm}

\section*{Acknowledgement}

Work supported by the GA\v{C}R grant Nr. 202/07/1307, by the
M\v{S}MT ``Doppler Institute" project Nr. LC06002  and by the
Institutional Research Plan AV0Z10480505.

\vspace{5mm}

\vspace{5mm}

\section*{Table captions}

\subsection*{Table 1. Auxiliary functions of couplings $\omega^{(J)}$
(even $N=2J$). }

\subsection*{Table 2. Auxiliary functions of couplings $\varepsilon^{(J)}$
(odd $N=2J+1$).}

\subsection*{Table 3. Coefficients
$C_{(n)}^{(J)}$ of eqs.~(\ref{definit}) and (\ref{pascal}). }

\subsection*{Table 4. Pascal-like  triangle, with coefficients
$C_{(n)}^{(J)}$ underlined }

\vspace{5mm}

\section*{Figure captions}

\subsection*{Figure 1. The $t-$dependence of energies,
with two choices of $A$ at $N=4$. }

\newpage


\begin{table}[t]
\caption{ Auxiliary functions of couplings $\omega^{(J)}$ (even
$N=2J$).} \label{pexp4}
\begin{center}
\begin{tabular}{||c|c||c||}
\hline \hline
  J&{\rm dimension}&
\multicolumn{1}{c||}{$\omega^{(J)}/[2\,(2J-1)\,(2J-1)!]$}\\
 \hline \hline
 2&4& A-B\\
 3&6& -3\,A+4\,B-C\\
 4&8&10\,A-15\,B+6\,C-D\\
 5&10&-35\,A+56\,B-28\,C+8\,D-E\\
 6&12&126\,A-210\,B+120\,C-45\,D+10\,E-F\\
 $\vdots$&$\vdots$&$\ldots$
 \\ \hline \hline
\end{tabular}
\end{center}
\end{table}


\begin{table}[b]
\caption{ Auxiliary functions of couplings $\varepsilon^{(J)}$
(odd $N=2J+1$). } \label{pexp1}
\begin{center}
\begin{tabular}{||c|c||c||}
\hline \hline
  J&{\rm dimension}&
\multicolumn{1}{c||}{$\varepsilon^{(J)}/[2\,(2J-1)\,(2J)!]$}\\
 \hline \hline
 2&5& A-B\\
 3&7& -2\,A+3\,B-C\\
 4&9&5\,A-9\,B+5\,C-D\\
 5&11&-14\,A+28\,B-20\,C+7\,D-E\\
 6&13&42\,A-90\,B+75\,C-35\,D+9\,E-F\\
 $\vdots$&$\vdots$&$\ldots$
 \\ \hline \hline
\end{tabular}
\end{center}
\end{table}

 \newpage

\begin{table}[t]
\caption{Coefficients $C_{(n)}^{(J)}$ of eqs.~(\ref{definit}) and
(\ref{pascal}). } \label{pexp2}
\begin{center}
\begin{tabular}{||c|cccccc|cccccc||}
\hline \hline
  &n=&
  $\ldots$&-3&-2&-1&0&1&2&3&4&5&
  $\ldots$\\
 \hline \hline
 J& & & & & &
 & &  &  &  &  &\\
 \hline
 \hline
 (1)  &&&{$\ldots$}&(0)&{(0)}&(-1)&(1)&{(0)}&(0)&{$\ldots$}&{}&\\
 2&&{}&{}&{}&-1&-1&1&1&{}&{}&{}&\\
 3&&{}&{}&-1&-3&-2&2&3&1&{}&{}&\\
 4&&{}&-1&-5&-9&-5&5&9&5&1&{}&\\
 5&&-1&-7&-20&-28&-14&14&28&20&7&1&\\
 $\vdots$ & & & & & &
  $\cdots$& $\cdots$&  &  &  &  &\\
  \hline \hline
\end{tabular}
\end{center}
\end{table}

\newpage

\begin{table}[b]
\caption{Pascal-like triangle, with coefficients $C_{(n)}^{(J)}$
underlined } \label{pexp3}
\begin{center}
\begin{tabular}{||c|ccccccccccccccc||}
\hline \hline
  J &&
  \multicolumn{5}{c}{ }&n&=&1&&2&&3&&$\ldots$\\
 \hline \hline
 (1)&&&&&&&-1&&{1}&&&&&&\\
 -&&&&{}&&-1&&0&&1&&&&&\\
 2&{}&&&&-1&&-1&&\underline{1}&&\underline{1}&&{}&&\\
 -&&&&-1&&-2&&0&&2&&1&&{}&\\
 3&&&-1&&-3&&-2&&\underline{2}&&\underline{3}&&\underline{1}&&{}\\
 -&&-1&&-4&&-5&&0&&5&&4&&1&\\
 4&-1&&-5&&-9&&-5&&\underline{5}&
 &\underline{9}&&\underline{5}&&\underline{1}\\
 $\vdots$&&&& & & &
  & & $\ldots$& & &&& &\\
  \hline \hline
\end{tabular}
\end{center}
\end{table}

\end{document}